\documentclass[pra,preprint,showpacs]{revtex4}
\usepackage[centertags]{amsmath}
\usepackage[koi8-r]{inputenc}
\usepackage{amsfonts}
\usepackage{amssymb}
\usepackage{amsthm} 
\usepackage{newlfont}
\usepackage{epsfig}
\usepackage{amscd}
\usepackage{graphicx}
\usepackage{epsfig}
\usepackage{footnote}
\usepackage{lipsum}
\usepackage{color}
\usepackage{xcolor}
\usepackage{graphicx}
\usepackage{multirow}
\usepackage{subfig}
\usepackage{amscd}

\newcommand{\beq}{\begin{equation}}
\newcommand{\eeq}{\end{equation}}
\newcommand{\ba}{\begin{array}}
\newcommand{\ea}{\end{array}}
\newcommand{\bea}{\begin{eqnarray}}
\newcommand{\eea}{\end{eqnarray}}
\begin{document}

\begin{center}
{\large \sc \bf {Complete structural restoring of transferred multi-qubit quantum state
}
}
\vskip 15pt

{\large 
E.B. Fel'dman$^{1,2}$,  A.N. Pechen$^{2,3}$ and  A.I.~Zenchuk$^{1,2}$ 
}

\vskip 8pt

{\it $^1$Institute of Problems of Chemical Physics, RAS,
Chernogolovka, Moscow reg., 142432, Russia}.

{\it $^2$Department of Mathematical Methods for Quantum Technologies, Steklov
Mathematical Institute of Russian Academy of Sciences, Gubkina str. 8,
Moscow
119991, Russia}

{\it $^3$National University of Science and Technology ''MISIS'', Leninski prosp. 4, Moscow 119049, Russia}

\vskip 8pt

\end{center}

\begin{abstract}
In this work, we study optimal transfer of quantum states via spin chains. We develop the protocol for structural restoring of multi-quantum coherence matrices of multi-qubit quantum states transferred from sender to receiver along a spin-1/2 chain. We also develop a protocol for constructing such  0-order coherence matrix that can be perfectly transferred in this process.
The restoring protocol is based on the specially constructed unitary transformation applied to the state of the extended receiver. This transformation for a given length of the chain {and Hamiltonian} is universally optimal in the sense that ones constructed it can be applied to optimally restore any higher-order coherence matrices.
\end{abstract}



\maketitle

\section{Introduction}
\label{Section:introduxtion}

The problem of quantum state transfer along the spin-1/2 chain was first formulated in the famous paper by Bose \cite{Bose}, where transfer of an arbitrary pure state along a homogeneous spin chain with Heisenberg Hamiltonian was considered.  In this case the state transfer was not perfect and fidelity~\cite{NC} averaged over pure initial states was used to estimate the deviation of the transferred state from the expected state. Similar to the teleportation problem   \cite{BBCJPW,BPMEWZ,BBMHP}, this subject attracted  great attention of researchers. A number of works were performed considering the ideal transfer~\cite{CDEL,KS} and high-probability state transfer along the inhomogeneous chains~\cite{GKMT}. Various aspects of state transfer were studied~\cite{FR,KF,CGIZ,GMT,DPF,BBVB}. Robustness to perturbations was widely investigated \cite{CRMF,ZASO,ZASO2,ZASO3} and it was shown that  chain which provides perfect  state transfer in the unperturbed case is equivalent to chain  providing high-probability state transfer for perturbed Hamiltonian~\cite{ZASO2}. Later the remote state transfer  method was introduced as an alternative to the state transfer. Initially this method was proposed for  photon systems  \cite{PBGWK,PBGWK2,DLMRKBPVZBW,PSB}, and then for  spin systems~\cite{LH,Z_2014,BZ_2015}. {In Refs. \cite{Z_2014,BZ_2015}, the extended receiver (i.e., a subsystem   including the receiver and its few nearest sites)  subjected to specially constructed unitary transformation was introduced for controlling the receiver's state in the state-creation process. In Refs.~\cite{FZ_2017,BFZ_Arch2018,Z_2018} a method of optimal transfer of quantum states via spin-1/2 chains was considered, where optimal transfer is understood as quantum state transfer with minimal and well characterized deformation of the transferred density matrix.}

{
Our paper is  further development of  the method underlined in Refs.~\cite{FZ_2017,BFZ_Arch2018,Z_2018}. {Before exploring the novelty of our paper, we briefly  summarize the results of Refs.~\cite{FZ_2017,BFZ_Arch2018,Z_2018}. First of all, 
recall} that   the density matrix  $ \rho$ of an $N$-qubit state of a spin system can always be written in the form:
\begin{eqnarray}
\rho=\sum_{-N}^N \rho^{(n)}.
\end{eqnarray}
Here $n$-order coherence matrix  $\rho^{(n)}$ includes  elements of the density matrix $\rho$ which correspond to state transitions changing the total spin $z$-projection by  $n$.
{It is shown in Ref. \cite{FZ_2017} that evolution of a spin chain under the Hamiltonian conserving the excitation number in the system (for instance, for nuclear spins in the strong magnetic field, the number of excited spins is the number of spins directed opposite the magnetic field; the Hamiltonian of the dipole-dipole interaction in the strong magnetic field conserves this number \cite{Abragam})  does not mix the coherence matrices of different orders $\rho^{(n)}$. This fact stimulated studying the state  transfer from 
 the sender to the receiver without such mixing in \cite{BFZ_Arch2018,Z_2018}.}

{In Refs.~\cite{BFZ_Arch2018,Z_2018} the sender ($S$) interacts with the receiver ($R$) through a spin chain called the transmission line ($TL$). However the initial state of the subsystem $TL\cup R$ is different in those works: it is the thermodynamic equilibrium state in \cite{BFZ_Arch2018} and ground state (i.e., the state without excited spins) in  \cite{Z_2018}. }

In Ref.  \cite{BFZ_Arch2018}, the block-scalable and block-scaled quantum states were found. A special property of these states is that { the receiver's state registered at some time instant $t_0$ differs from the sender's initial state  by multiplication of the $n$-order ($|n|>0$)  coherence matrices by certain scale factors $\lambda^{(n)}$}.
%
But the 0-order coherence matrix must satisfy the trace-normalization condition and therefore can not be completely scaled. However,  in contrast to non-zero-order coherence matrices, the 0-order coherence matrix can be perfectly transferred, i.e.,  transfer with $\lambda^{(0)}=1$ is possible.  {We emphasize that in Refs.~\cite{FZ_2017,BFZ_Arch2018} the extended receiver with optimizing unitary transformation was not involved into the control process.}   

{ As was shown in~\cite{BFZ_Arch2018}, for block-scaled state transfer} each pair of $\pm n$-order coherence matrices  can transfer not more than one arbitrary  complex parameter. Therefore the information capacity of the block-scaled state transfer is small. This makes important finding its modifications to increase the information capacity and motivates development of protocols for restoring the structure of the non-diagonal part of the density matrix of the transferred state using special unitary transformation of the extended receiver so that  the non-diagonal elements of the receiver state would differ only by fixed scale factors from the corresponding elements of the initial sender's state. Then each non-diagonal element would transfer its own parameter. An example of such protocol for restoring a 2-qubit state was proposed in 
Ref.~\cite{Z_2018}. But the method of unitary transformations does not work for restoring diagonal elements because they have different structures. Therefore, the problem of restoring the diagonal elements was completely disregarded in~\cite{Z_2018}.

 
{ The goal of our paper is two-fold. First,  we extend the 2-qubit structural restoring protocol of Ref.~\cite{Z_2018} and, using the tool of multi-index technique, derive general formulas for restoring non-diagonal elements  of multi-qubit quantum states using the  optimizing unitary transformation of the extended receiver. Second, we explore the method of manipulating with the elements of 0-order coherence matrix developing the idea of \cite{BFZ_Arch2018}. Similar to Ref.~\cite{Z_2018}, the subsystem $TL\cup R$ is in the ground initial state without excitations. This general theoretical analysis is then applied to state-restoring protocol for a 2-qubit state evolved under $XX$-Hamiltonian which, together with other aspects of a 2-qubit state restoring given in \cite{BFZ_Arch2018,Z_2018}, provides a comprehensive description of the state-restoring process.}

 We emphasize that the scale factors { found in  Ref. \cite{BFZ_Arch2018}, the scale factors together with 
the optimizing unitary transformation of the extended receiver found in Ref.~\cite{Z_2018} and  in the present work}  are universal objects in the sense that they are defined only by  the evolution Hamiltonian (in particular, by the chain length), { by the initial state of the subsystem $TL\cup R$} and by the  time instance fixed for the receiver state registration {and do not depend on} the particular initial state of the sender {to be transferred. Thus, the registered receiver's state depends on the initial sender's state, but the scale factors obtained in the restoring process and the appropriate unitary transformation do not depend on the initial state of sender}. Hence once constructed, this unitary transformation can be applied to restore non-diagonal elements of any transferred { sender's initial} state. 
This universality notion is related to work~\cite{WuPe2007}, where the most general class of universally optimal Kraus maps was described and investigated for quantum control (Kraus maps are most general transformations of quantum states~\cite{Kraus}).}
{The state-restoring method developed below is applicable to any ground and thermodynamically equilibrium initial states of the subsystem $TL\cup R$. However, using the ground initial state allows us to 
reduce the dimension of the  Hilbert space of the whole $N$-qubit chain (whose dynamic is to be described) from $2^N$ to 
$\sum_{k=1}^{N^{(S)}} C^k_{N^{(S)}}$ (where $N^{(S)}$ is the number of qubits in the sender, $N^{(S)}<N$) and therefore simplifies calculations. This remark motivates our choice of the ground state of the subsystem $TL\cup R$.}
{It is important that  the restoring unitary transformation of the extended receiver is not unique \cite{Z_2018}, and it can be used for further optimization (for instance, maximization by the absolute values) of  the scale factors in the restorer state. We perform some rough optimization in this work. Global optimization techniques which were applied for controlling spins~\cite{PeRa2006,Morzhin2019,Morzhin2020,PPN2020,Volkov2021} could be used as well.}

{ We remark that the perfect (and almost perfect up to one diagonal element) state transfer was achieved in Ref.~\cite{BFZ_Arch2018}  by the method which is not applicable to the case of ground initial state of the subsystem $TL\cup R$. } {Therefore in this work for ground initial state  we develop a different method and show that there exists a special 0-order coherence matrix which can be almost perfectly transferred, up to two diagonal elements, which must satisfy the trace-normalization condition.
Furthermore, 
there exists such 0-order coherence matrix that can be perfectly transfered from the sender to the receiver up to the trivial exchange of two its diagonal elements. 
We call such transfer as perfect transfer of 0-order coherence matrix. 
}


The  paper is organized as follows.
Evolution of the sender's initial state and structure of the receiver's  state under the unitary transformation of the extended receiver is analyzed in Sec.~\ref{Section:evolution}.  General protocol for state restoring based on the optimizing unitary transformation of the extended 
receiver is developed in Sec.~\ref{Section:restoring}. In the same section, we also find a 0-order coherence matrix which can be perfectly transferred to the receiver. An example of a two-qubit state restoring is constructed in Sec.~\ref{Section:example}. Conclusions are provided in Sec.~\ref{Section:conclusion}. Some additional important formulas are given in the Appendix Sec.~\ref{Section:appendix}.

\section{Evolution operator and unitary transformation of the extended receiver}
\label{Section:evolution}

\begin{figure*}[!]
\epsfig{file=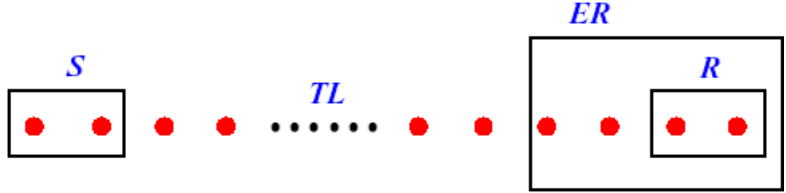,
  scale=0.6
   ,angle=0
} 
\caption{Communication line consisting of sender S, transmission line TL and receiver R. 
The extended receiver ER is shown as a box which includes the receiver R and its several nearest nodes.  
}
\label{Fig:rec} 
\end{figure*}

We consider the communication line shown in Fig.~\ref{Fig:rec} and consisting of the sender $S$, where the state to be transferred is initiated, the transmission line $TL$, along which the state is transferred from the sender to the receiver, and the receiver $R$, where the transferred state is registered. 
We also will use an extended receiver $ER$~\cite{Z_2018} which is necessary to handle in a desired manner the state of the receiver via controlled unitary transformation. The receiver is a part of the extended receiver. By $\overline{TL}$ denote the transmission line without nodes of the extended receiver. The sender and the receiver are considered to have the same dimension:{
\begin{eqnarray}\label{NSNR}
N^{(S)} = N^{(R)}\equiv N.
\end{eqnarray}}

Evolution of the density matrix of the spin chain is given by unitary transformation with Hamiltonian $H$,
\begin{eqnarray}\label{erv}
\rho(t) = V(t) \rho(0) V^\dagger(t),\qquad V(t)=e^{-i H t},
\end{eqnarray}
To prevent mixing of the { multi-quantum  coherence matrices of different orders}  we require that the Hamiltonian $H$  satisfies the commutation condition \cite{FZ_2017}
\begin{eqnarray}\label{com}
[H,I_z]=0\, ,
\end{eqnarray}
 and that the initial state $\rho(0)$ has tensor-product form,
\begin{eqnarray}\label{incon}\label{InStS}
\rho(0)=\rho^{(S)}(0)\otimes \rho^{(TL,R)}(0).
\end{eqnarray}
Here $\rho^{(S)}(0)$ is the sender's initial density matrix whose structure will be discussed below, and  $\rho^{(TL,R)}(0)$ is the initial density matrix of the subsystem $TL\cup R$, {which is the ground state:
\begin{eqnarray}\label{InStTLR}
\rho^{(TL,R)}(0)={\mbox{diag}}(1,0,\dots,0)\, .
\end{eqnarray}}
It is important that  $\rho^{(TL,R)}(0)$ contains only the 0-order coherence matrix. This condition together with the commutation condition (\ref{com}) provides  transfer  of a quantum state  from the sender to the receiver without interaction among multi-quantum  coherence matrices  \cite{FZ_2017}. 

At some time instant $t_0$ we apply to the extended receiver a unitary transformation $U$ which depends on the set of free parameters $\varphi$. { This transformation is required to conserve} the excitation number, similarly to $H$, i.e.,  
\begin{eqnarray}\label{st}
[I^{ER}_z,U(\varphi)]=0,
\end{eqnarray}
where $I^{ER}_z$ is the $z$-projection of the total spin moment in the state-space of the extended receiver $ER$.
Then the state of the system becomes
\begin{eqnarray}\label{rtv}
\rho(t_0,\varphi)=W(t_0,\varphi) \rho(0) W^\dagger(t_0,\varphi),
\end{eqnarray}
where
\begin{eqnarray}
W(t_0,\varphi)=(\mathbb I_{S,\overline{TL}}\otimes U(\varphi)) V(t_0)\,.
\end{eqnarray}
Here $\mathbb I_{S,\overline{TL}}$ is the identity operator in the subspace of the subsystem $S\cup \overline{TL}$. {Notice that the dynamics of the initial state (\ref{incon}) with $N$-qubit sender can be described in the subspace of up to $N$ excitations due to the commutation relation $[W,I_z]=0$ {(which follows from (\ref{com}) and (\ref{st}))  and the initial state (\ref{InStS}), (\ref{InStTLR}).} Therefore both operators $U$ and  $W$ should have the following  block-diagonal form:}
\begin{eqnarray}\label{U}
U(\varphi)&=&{\mbox{diag}}(1, U^{(1)}(\varphi^{(1)}),\dots,U^{(N)}(\varphi^{(N)})),\\
\label{W}
W&=&{\mbox{diag}}({1}, W^{(1)}(\varphi^{(1)}),\dots,W^{(N)}(\varphi^{(N)})),\\\nonumber
\varphi&=&(\varphi^{(1)},\dots,\varphi^{(N)}).
\end{eqnarray}
where $W^{(k)}$ is  the block in the subspace of $k$ excited spins and $\varphi^{(k)}$ is the list of free parameters in this block, $k=0,\dots, N$. 

The dimension of the $k$th block,
\begin{eqnarray}\label{UC}
C^{k}_{N} \times C^{k}_{N},\qquad C^{k}_{N} =\frac{N!}{k! (N-k)!},
\end{eqnarray}
defines the number of free real parameters  in the block $U^{(k)}$:
\begin{eqnarray}\label{Uvarphi}
\varphi^{(k)} = (\varphi^{(k)}_1,\dots,\varphi^{(k)}_{C^{k}_{N}(C^{k}_{N}-1)}).
\end{eqnarray}
State of the receiver at the time instant $t_0$ can be found from $\rho(t_0,\varphi)$ by tracing over $S$ and $TL$:  
\begin{eqnarray}\label{RTr}
\rho^{(R)}(t_0,\varphi)={\mbox{Tr}}_{S,TL}(\rho(t_0,\varphi)).
\end{eqnarray}
Hereafter for shortness of notations we will sometimes omit the parameters $t_0$ and $\varphi$ from the arguments.

\subsection{Structure of the transformation (\ref{rtv})}
Now we proceed to study the structure of the transformation~(\ref{rtv}).
Before writing this equation in components, we introduce the multi-indexes for the subsystems $S$, $TL$, and $R$, which we denote by the capital latin letters with the corresponding  subscripts. For instance, $I_S$ is the multi-index linked to the sender  $S$. The multi-index is the set of zeros and ones, the cardinality of this set equals to the number of qubits in the subsystem. The number 1 at a certain position indicates that the corresponding spin is excited, while the number 0 indicates that the spin is in the ground state. We also introduce the norm of the multi-index $|\cdot|$, which is the number of excited spins in the subsystem (and equals to the sum of all elements in this multi-index). The notation of indexed 1 and 0 {(for instance, $1_S$, $0_R$)} are used to indicate the indexes with all ones or zeros. Since both sender and receiver have the same dimension {(\ref{NSNR})}, the multi-indexes with subscripts $R$ and $S$ are of equal length, so that in particular,  $1_R = 1_S$ and $0_R=0_S$. In these terms, the scalar blocks $W^{(0)}$ and $ W^{(N)}$ of the unitary transformation $W$ given by Eq. (\ref{W}) have the form:
\begin{eqnarray}\label{W0}
&&W^{(0)}=W^{(0)}_{0_S0_{TL}0_R;0_S0_{TL}0_R} =1,\\\label{W1}
&&W^{(N)} = W^{(N)}_{1_S1_{TL}1_R;1_S1_{TL}1_R}.
\end{eqnarray}
In general, elements of the $n$th block $
W^{(n)}_{I_SI_{TL}I_R;J_SJ_{TL}J_R} 
$ satisfy
\begin{eqnarray}\label{ststst}
|I_S|+|I_{TL}|+|I_R|=|J_S|+|J_{TL}|+|J_R|=n.
\end{eqnarray}
Now the density matrix $\rho(t_0,\varphi)$ (\ref{rtv}) can be written in the following component form:
\begin{eqnarray}\label{rhoNM}
\rho_{N_SN_{TL}N_R;M_SM_{TL}M_R} = \sum_{I_S,J_S} W^{(|I_S|)}_{N_SN_{TL}N_R;I_S 0_{TL} 0_R}  \rho^{(S)}_{I_S,J_S} 
(W^{(|J_S|)})^\dagger_{J_S 0_{TL} 0_R;M_SM_{TL}M_R} .
\end{eqnarray}

\subsection{Structure of the receiver's state (\ref{RTr})}

Tracing over  $S$ and $TL$ gives the receiver's density matrix $\rho^{(R)}$:
\begin{eqnarray}\label{rhoNMR0}\label{rhoNMR}
\rho^{(R)}_{N_R,M_R} &=& \sum_{{N_S,N_{TL}}\atop{|N_S|+|N_{TL}|=|I_S|-|N_R|}} \sum_{I_S,J_S}
W^{(|I_S|)}_{N_SN_{TL}N_R;I_S 0_{TL} 0_R}  \rho^{(S)}_{I_S,J_S} 
(W^{(|J_S|)})^\dagger_{J_S 0_{TL} 0_R;N_SN_{TL}M_R}.
\end{eqnarray}
Expression~(\ref{rhoNMR0}) can be written in the matrix form as follows:
\begin{eqnarray}\label{rhoNMR0g}
\rho^{(R)} &=& \sum_{N_S,N_{TL}}
\hat W_{N_SN_{TL}}  \rho^{(S)} 
\hat W^\dagger_{N_SN_{TL}} ,
\end{eqnarray}
where $\hat W_{N_SN_{TL}}$ are Kraus operators~\cite{Kraus} which satisfy
\begin{eqnarray}\label{kraus}
\sum_{N_S,N_{TL}}\hat W^\dagger_{N_SN_{TL}} \hat W_{N_SN_{TL}} =\mathbb I_{S},
\end{eqnarray}
and have elements
\begin{eqnarray}
(\hat W_{N_SN_{TL}})_{N_R;I_S}=\left\{
\begin{array}{ll}
W^{(I_S)}_{N_SN_{TL}N_R;I_S0_{TL}0_R},& |N_S|+|N_{TL}|+|N_R|=|I_S|\cr
0,& |N_S|+|N_{TL}|+|N_R|\neq |I_S|
\end{array}
\right.,
\end{eqnarray}
Here $\mathbb I_{S}$ is the identity operator in the subspace of $S$. In (\ref{rhoNMR0g}) and (\ref{kraus}), the sum is over all $N_S$ and $N_{TL}$.

{Now we prove that there is no mixing of elements of different   multi-quantum coherence matrices of the matrix  $\rho^{(S)}(0)$}
in  (\ref{rhoNMR0}).
In fact, by virtue of (\ref{ststst}) one has that~(\ref{rhoNMR})
\begin{eqnarray}\label{NNN}
|N_S|+|N_{TL}|+|N_R| =|I_S|,\qquad |N_S|+|N_{TL}|+|M_R| =|J_S|.
\end{eqnarray}
Since by definition {
the elements  $\rho^{(S)}_{I_S,J_S}$ and $\rho^{(R)}_{N_R,M_R}$  are included in, respectively, the coherence matrices of the order  $n=|J_S|-|I_S|$
and
$m=|M_R|-|N_R|$,} taking into account  (\ref{NNN}) gives:
\begin{eqnarray}
m=|M_R|-|N_R| = (|J_S| -|N_S|-|N_{TL}|) - (|I_S|-|N_S|-|N_{TL}|)= |J_S|-|I_S| =n.
\end{eqnarray}
This means that elements of  the $n$-order coherence matrix of the sender 
become equal to elements of the $n$-order coherence matrix of the receiver.
{Then, using the expansion of  $\rho^{(S)}$ and $\rho^{(R)}$ in terms of the  multi-quantum coherence matrices $\rho^{(S;n)}$ and $\rho^{(S;n)}$,
\begin{eqnarray}\label{stst}
\rho^{(S)}&=&\sum_{-N^{(S)}}^{N^{(S)}} \rho^{(S;n)}\, ,
\\\label{nR}
\rho^{(R)}&=&\sum_{-N^{(S)}}^{N^{(S)}} \rho^{(R;n)}\, ,
\end{eqnarray}}
  Eq.~(\ref{rhoNMR}) can be written as
\begin{eqnarray}\label{rhoNMRf}
&&
\rho^{(R;n)}_{N_R,M_R}  =\\\nonumber
&&\sum_{{N_S,N_{TL}}\atop{|N_S|+|N_{TL}|=|I_S|-|N_R|}}
\sum_{I_S,J_S}
W^{(I_S)}_{N_SN_{TL}N_R;I_S 0_{TL} 0_R}   \rho^{(S;n)}_{I_S,J_S}
(W^{(J_S)})^\dagger_{J_S 0_{TL} 0_R;N_SN_{TL}M_R}
, \\\nonumber
&&
-N\le n \le N.
\end{eqnarray}

\subsection{Multi-quantum coherence matrices}

Now we consider the structure of the  multi-quantum coherence matrices of positive order. The coherence matrices of negative order are Hermitian conjugate of {these matrices}. 0-order and higher order coherence matrices play different roles in state transfer. Only higher-order coherence matrices of the sender's initial state  can be effectively restored at the receiver's state and thus  serve to transfer arbitrary parameters from the sender to the receiver.  On the contrary, sender's  0-order coherence matrix
can not be completely restored  at the receiver side. But this matrix can be fixed  in such a way that most of its elements (or even all of them) are perfectly transferred to the receiver.
In this way we provide the minimal deformation of the sender's initial state during its transfer to the receiver. 
 
\subsubsection{Higher-order coherence matrices}
\paragraph{$N$-order coherence matrix.}
\label{Section:Nth}
We select the $N$-order coherence matrix because the condition $|M_R|-|N_R|=N$ can be satisfied only if 
 $N_R=0_R$, $M_R=1_R$, i.e., when there is only one element in this matrix. Eq.~(\ref{rhoNMRf}) yields { in view of (\ref{W0})}
\begin{eqnarray}\label{rhoNMR0h}
\rho^{(R;N)}_{0_R,1_R} &=&
W^{(0)}_{0_S0_{TL}0_R;0_S 0_{TL} 0_R}  \rho^{(S;N)}_{0_S,1_S} 
(W^{(N)})^\dagger_{1_S 0_{TL} 0_R;0_S0_{TL}1_R}\\\nonumber
&=&\rho^{(S;N)}_{0_S,1_S} 
(W^{(N)})^\dagger_{1_S 0_{TL} 0_R;0_S0_{TL}1_R} =\rho^{(S;N)}_{0_S,1_S} 
W^{(N)}_{0_S0_{TL}1_R;1_S 0_{TL} 0_R} .
\end{eqnarray}

{
Following Ref.~\cite{Z_2018}, we use the intensity of this coherence to fix the {optimal}  time instant $t_0$ for receiver's state registration. Namely, we take the time instant $t_0$ corresponding to the maximal absolute value of the scale factor in $\rho^{(R;N)}_{0_R,1_R}$ with zero values of all parameters $\varphi$}:
\begin{eqnarray}\label{t_0}
\max_t
|W^{(N)}_{0_S0_{TL}1_R;1_S 0_{TL} 0_R}|_{\varphi=0} = 
|W^{(N)}_{0_S0_{TL}1_R;1_S 0_{TL} 0_R}|_{{\varphi=0}\atop{t=t_0}} .
\end{eqnarray}

\paragraph{$k$-order coherence matrix, $1\le k<N$.}
\label{Section:kth}
For matrix elements $\rho^{(R;k)}_{N_R,M_R}$ of the $k$-order coherence matrix one has 
$|M_R|-|N_R|=k$. First, select the following family of elements:
$|M_R|=N$ (i.e., $M_R=1_R$) and $|N_R| = N-k$. For these elements, one can write
\begin{eqnarray}\label{rhoNMR0g0}
\rho^{(R;k)}_{N_R,1_R} = 
\sum_{|I_S|=N-k} W^{(N-k)}_{0_S0_{TL}N_R;I_S0_{TL}0_R} \rho^{(S;k)}_{I_S,1_S} 
(W^{(N)})^\dagger_{1_S 0_{TL} 0_R; 0_S 0_{TL} 1_R}.
\end{eqnarray}
For other elements, one has  $0<|M_R|<N$, and one can write the general formula (which reduces to (\ref{rhoNMR0g0}) for $|M_R|=N$):
\begin{eqnarray}\label{rhoNMR0g2}
\rho^{(R;k)}_{N_R,M_R} &=& \\\nonumber
&&\sum_{{N_S,N_{TL}}\atop{|N_S|+|N_{TL}|=|I_S|-|N_R|}}
\sum_{{I_S,J_S, |I_S|\ge N_R}\atop{|J_S|-|I_S|=k}} 
W^{(|I_S|}_{N_SN_{TL}N_R;I_S 0_{TL} 0_R}  \rho^{(S;k)}_{I_S,J_S} 
(W^{(|J_S|})^\dagger_{J_S 0_{TL} 0_R;N_SN_{TL} M_R} .
\end{eqnarray}

\subsubsection{0-order coherence matrix}
\label{Section:zer0}
For 0-order coherence matrix one has in (\ref{rhoNMR0}):
\begin{eqnarray}
|N_R|=|M_R|,\qquad |I_S|=|J_S|.
\end{eqnarray}
Then the matrix elements have the following representation:
\begin{eqnarray}\label{rhoNMR02}\label{rhoNMR0c}
&&
\rho^{(R;0)}_{N_R,M_R} = \\\nonumber
&&\sum_{{N_S,N_{TL}}\atop{|N_S|+|N_{TL}|=|I_S|-|N_R|}}
\sum_{{I_S,J_S}\atop{|I_S|=|J_S|\ge |N_R|}}
W^{(|I_S|)}_{N_SN_{TL}N_R;I_S 0_{TL} 0_R}  \rho^{(S;0)}_{I_S,J_S} 
(W^{(|J_S|)})^\dagger_{J_S 0_{TL} 0_R;N_SN_{TL}M_R} .
\end{eqnarray}
To check the normalization condition, we set $M_R=N_R$ and calculate the sum of  eqs.~(\ref{rhoNMR02}) over $N_R$ to obtain:
\begin{eqnarray}\label{norm}
\sum_{N_R}\rho^{(R;0)}_{N_R,N_R}=\sum_{I_S} \rho^{(S;0)}_{I_S,I_S} =1.
\end{eqnarray}

We analyze the structure of Eq.~(\ref{rhoNMR02}) for the two particular cases. 

First, 
let $|N_R|=|M_R|=N$. This is possible only if $N_R=M_R=1_R$. Then (\ref{rhoNMR02}) yields
\begin{eqnarray}\label{rhoNMR0a}
\rho^{(R;0)}_{1_R,1_R} &=& 
W^{(N)}_{0_S 0_{TL}1_R;1_S 0_{TL} 0_R}  \rho^{(S;0)}_{1_S,1_S} 
(W^{(N)})^\dagger_{1_S 0_{TL} 0_R;0_S 0_{TL}1_R}\\\nonumber
&=&
|W^{(N)}_{0_S 0_{TL}1_R;1_S 0_{TL} 0_R}|^2  \rho^{(S;0)}_{1_S,1_S} ,
\end{eqnarray}
i.e., the element $\rho^{(R;0)}_{1_R,1_R}$ is proportional to $\rho^{(S;0)}_{1_S,1_S}$.

Second, let$|N_R|=|M_R|=0$. Then $N_R=M_R=0_R$ and Eq.~(\ref{rhoNMR0c}) yields
\begin{eqnarray}\label{rhoNMR0d}
\rho^{(R;0)}_{0_R,0_R} &=&
\\\nonumber
&&\sum_{{N_S,N_{TL}}\atop{|N_S|+|N_{TL}|=|I_S|-|N_R|}}
\sum_{{I_S,J_S}\atop{0\le|I_S|=|J_S|\le N}}
W^{(|I_S|)}_{N_SN_{TL}0_R;I_S 0_{TL} 0_R}  \rho^{(S;0)}_{I_S,J_S} 
(W^{(|J_S|)})^\dagger_{J_S 0_{TL} 0_R;N_S N_{TL}0_R} .
\end{eqnarray}

{\bf Remark.} If all elements of $\rho^{(S;0)}$ equal to zero except the single one 
$\rho^{(S;0)}_{0_S,0_S}=1$, i.e.,
\begin{eqnarray}\label{rho00}
\rho^{(S;0)}_{N_SM_S} = \delta_{N_S,0_S} \delta_{M_S,0_S},  
\end{eqnarray}
where $\delta$ is the Kronecker symbol, then
Eq.~(\ref{rhoNMR02}) reduces to the following one
\begin{eqnarray}\label{rhoNMR03}
\rho^{(R;0)}_{N_R,M_R} &=& 
\\\nonumber
&&
W^{(0)}_{0_S0_{TL}N_R;0_S 0_{TL} 0_R}  \rho^{(S;0)}_{0_S,0_S} 
(W^{(0)})^\dagger_{0_S 0_{TL} 0_R;0_S0_{TL}M_R} = \delta_{N_R, 0_R}\delta_{M_R,0_R}\rho^{(S;0)}_{0_R,0_R} ,
\end{eqnarray}
where we take into account that $W_{0_S0_{TL}0_R;0_S 0_{TL} 0_R}\equiv 1$.
Thus, if the 0-order coherence matrix of the sender's initial state is 
\begin{eqnarray}\label{rhosd}
\rho^{(S;0)}={\mbox{diag}} (1,0,\dots,0),
\end{eqnarray}
then this matrix is transfered to the state of the receiver without any change, so that 
$\rho^{(R;0)} \equiv \rho^{(S;0)} \equiv {\mbox{diag}} (1,0,\dots,0)$.
However, any correction to  (\ref{rhosd}) destroys the positivity  of the density matrix and therefore it  can not be used as a 0-order coherence matrix of the sender's initial state for the purpose of  transferring the  higher order coherence matrices to the receiver.

\section{General state restoring protocol}
\label{Section:restoring}
In summary, the restoring of the sender's initial state at the receiver includes three steps.
\begin{enumerate}
\item
Fixing the time instant for the receiver's state registration.
\item
Constructing the optimizing unitary transformation of the extended receiver.
\item
Constructing the  0-order coherence matrix of special form in the sender's initial state.
\end{enumerate}

{The time instant $t_0$ was fixed in Sec.~\ref{Section:Nth} by Eq.~(\ref{t_0}). Two other points are explored below.}

\subsection{Unitary transformation of the extended receiver} 
As stated above, the unitary transformation of the extended receiver is needed to  obtain density matrix $\rho^{(R)}$ at some time instant such that elements of its higher order coherence matrices satisfy the condition
\begin{eqnarray}\label{trSRn}
\rho^{(R;n)}_{N_R,M_R} = \lambda^{(n)}_{N_R,M_R} \rho^{(S;n)}_{N_R,M_R},\qquad |n|>0\,.
\end{eqnarray}
We remind that the subscripts $S$ and $R$ are equivalent in multi-index notation. Below we study this condition in detail.

\paragraph{$N$-order coherence matrix.}
According to the definition (\ref{trSRn}) and formulae of Sec.\ref{Section:Nth},
the $N$-order coherence matrix has one element and therefore does not require restoring:
\begin{eqnarray}\label{N2cohlam1}
\lambda^{(N)}_{0_R,1_R}=
(W^{(N)})^\dagger_{1_S 0_{TL} 0_R;0_S0_{TL}1_R}=
W^{(N)}_{0_S0_{TL}1_R;1_S 0_{TL} 0_R}.
\end{eqnarray}
\paragraph{$k$-order coherence matrices, $1\le k < N$.}
Eq.~(\ref{rhoNMR0g0}) yields
\begin{eqnarray}\label{kcohonstr0}
W^{(N-k)}_{0_S0_{TL}N_R;I_S0_{TL}0_R} 
(W^{(N)})^\dagger_{1_S 0_{TL} 0_R; 0_S 0_{TL} 1_R}= \lambda^{(k)}_{N_R1_R} \delta_{N_R,I_S},\qquad |N_R|=|I_S| = N-k.
\end{eqnarray}
Eq.~(\ref{rhoNMR0g2}) yields
\begin{eqnarray}\label{kcohonstr}
&&\sum_{{N_S,N_{TL}}\atop{|N_S|+|N_{TL}|=|I_S|- |N_R|}}
W^{(|I_S|}_{N_SN_{TL}N_R;I_S 0_{TL} 0_R}  
(W^{(|J_S|})^\dagger_{J_S 0_{TL} 0_R;N_SN_{TL} M_R}  =\lambda^{(k)}_{N_RM_R} \delta_{N_R,I_S} \delta_{M_R,J_S},\\\nonumber
&&|J_S|-|I_S| =|M_R|-|N_R| = k, \qquad 0<|M_R|<N.
\end{eqnarray}
Eq.~(\ref{kcohonstr0}) with zero rhs  yields the constraint
\begin{eqnarray}\label{kconstr1}
W^{(N-k)}_{0_S0_{TL}N_R;I_S 0_{TL} 0_R}=0 ,\qquad N_R\neq I_S,\qquad |N_R|=|I_S| = N-k.
\end{eqnarray}
Eq.~(\ref{kcohonstr0}) with non-zero rhs results in  the following expressions for the scale factors:
\begin{eqnarray}\label{klambda1}
\lambda^{(k)}_{N_R1_R} = 
W^{(N-k)}_{0_S0_{TL}N_R;N_R 0_{TL} 0_R} (W^{(N)})^\dagger_{1_S0_{TL}0_R; 0_S0_{TL}1_R},\qquad |N_R|=N-k.
\end{eqnarray}
Eq.~(\ref{kcohonstr})  with zero rhs yields the additional constraints
\begin{eqnarray}\label{kconstr2}
&&
\sum_{{N_S,N_{TL}}\atop{|N_S|+|N_{TL}|=|I_S|- |N_R|}}
W^{(|I_S|}_{N_SN_{TL}N_R;I_S 0_{TL} 0_R} 
(W^{(|J_S|})^\dagger_{J_S 0_{TL} 0_R;N_SN_{TL} M_R}  =0,\\\nonumber
&&
N_R\neq I_S \qquad {\mbox{or}} \qquad
M_R\neq J_S,\qquad 0<|M_R|<N,  \qquad |J_S|-|I_S| = |M_R|-|N_R|=k.
\end{eqnarray}
 Eq.(\ref{kcohonstr})  with non-zero rhs results in  the following expressions for the scale factors:
\begin{eqnarray}\label{klambda2}
\lambda^{(k)}_{N_RM_R}&=&W^{(|N_R|}_{0_S 0_{TL}N_R;N_R 0_{TL} 0_R}  
(W^{(|M_R|})^\dagger_{M_R 0_{TL} 0_R;0_S 0_{TL} M_R}, \\\nonumber
&& |M_R|-|N_R|=k,\qquad 0<|M_R|<N. 
\end{eqnarray}

\paragraph{Optimization of the scale factors.}
The parameters $\varphi$ of the unitary transformation must satisfy the constraints 
(\ref{kconstr1}) and (\ref{kconstr2}).
We call the  unitary transformation $U$ with parameters satisfying these  constraints as the optimizing unitary transformation $U_{opt}$ of the extended receiver.  In addition, we also can use the parameters $\varphi$ to optimize (for instance, to maximize the absolute values) 
the scale factors given in  expressions (\ref{N2cohlam1}),  
(\ref{klambda1}) and (\ref{klambda2}).
The number of the parameters $\varphi$ must be large enough to satisfy all the above constraints and requirements. This number depends on the dimension of the unitary transformation which in turn is defined by the dimension of the extended receiver.

\subsection{Construction of the 0-order coherence matrix}
\subsubsection{Elements which can be perfectly transferred and the normalization condition}
\label{Section:0orderNorm}
In this section we find the structure of the 0-order coherence matrix, which can be either nearly perfectly transferred (up to two elements included into the blocks { of 0 and $N$ excitations}) or perfectly transferred.

First of all one has to take into account constraints 
(\ref{kconstr1}) and (\ref{kconstr2}) 
obtained in restoring the higher order coherence matrices.
Eq.~(\ref{rhoNMR0a}) for $|N_R|=|M_R|=N$ remains the same. 
Eq.~(\ref{rhoNMR0c}) for $0<|N_R|=|M_R|<N$ yields
\begin{eqnarray}\label{rhoNMR0cC}
\rho^{(R;0)}_{N_R,M_R} &=&\\\nonumber
&&
W^{(|N_R|)}_{0_S0_{TL}N_R;N_R 0_{TL} 0_R}  \rho^{(S;0)}_{N_R,M_R} 
(W^{(|M_R|)})^\dagger_{M_R 0_{TL} 0_R;0_S0_{TL}M_R}\\\nonumber
&&+\sum_{{N_S,N_{TL}}\atop{|N_S|+|N_{TL}|=|I_S|-|N_R|}}
\sum_{{I_S,J_S}\atop{|N_R|<|I_S|=|J_S|<N}}
W^{(|I_S|)}_{N_SN_{TL}N_R;I_S 0_{TL} 0_R}  \rho^{(S;0)}_{I_S,J_S} 
(W^{(|J_S|)})^\dagger_{J_S 0_{TL} 0_R;N_SN_{TL}M_R}\\\nonumber
&&+\sum_{{N_S,N_{TL}}\atop{|N_S|+|N_{TL}| = N-|N_R|}}
W^{(N)}_{N_SN_{TL}N_R;1_R 0_{TL} 0_R}  \rho^{(S;0)}_{1_R,1_R} 
(W^{(N)})^\dagger_{1_R 0_{TL} 0_R;N_SN_{TL}M_R}
\end{eqnarray}
Finally, Eq.~(\ref{rhoNMR0d})  for $|N_R|=|M_R|=0$ reads
\begin{eqnarray}\label{rhoNMR0dC}
\rho^{(R;0)}_{0_R,0_R} &=&\rho^{(S;0)}_{0_R,0_R}
\\\nonumber
&=&
\sum_{{N_S,N_{TL}}\atop{|N_S|+|N_{TL}| = |I_S|}}
\sum_{{I_S,J_S}\atop{0<|I_S|=|J_S|<N}}
W^{(|I_S|)}_{N_SN_{TL}0_R;I_S 0_{TL} 0_R}  \rho^{(S;0)}_{I_S,J_S} 
(W^{(|J_S|)})^\dagger_{J_S 0_{TL} 0_R;N_S N_{TL}0_R}
\\\nonumber
&&+
\sum_{{N_S,N_{TL}}\atop{|N_S|+|N_{TL}|=N}}
|W^{(N)}_{N_SN_{TL}0_R;1_R 0_{TL} 0_R} |^2 \rho^{(S;0)}_{1_R,1_R} 
.
\end{eqnarray}

Now we look for such elements which can be perfectly transferred to the receiver. For the element $\rho^{(R;0)}_{1_R,1_R}$ given in (\ref{rhoNMR0a})  this requirement $\rho^{(R;0)}_{1_R,1_R}=\rho^{(S;0)}_{1_R,1_R}$ may not be satisfied for nonzero $\rho^{(S;0)}_{1_R,1_R}$, because $|W^{(N)}_{0_S 0_{TL}1_R;1_S 0_{TL} 0_R}| < 1 $ for $t>0$. Therefore this element can not be perfectly transferred. There must be at least one more non-perfectly transferred diagonal element to provide the normalization
{\begin{eqnarray}\label{norm0}
\sum_{N_R}\rho^{(R;0)}_{N_R,N_R}=1.
\end{eqnarray}}
Let $\rho^{(S;0)}_{0_R,0_R}$ be such element. 

The perfect transfer of the elements $\rho^{(S;0)}_{N_R;M_R}$ with $0<|N_R|=|M_R|<N$ can be arranged using
Eq.~(\ref{rhoNMR0cC}). 
Setting 
$\rho^{(R;0)}_{N_R;M_R}=\rho^{(S;0)}_{N_R;M_R}$ in this system, we obtain in this case the following equation for $\rho^{(S;0)}_{N_R,M_R}$:
\begin{eqnarray}\label{rhoNMR0bz}
\rho^{(R;0)}_{N_R,M_R} &=&\rho^{(S;0)}_{N_R,M_R} \\\nonumber
&=&
W^{(|N_R|)}_{0_S0_{TL}N_R;N_R 0_{TL} 0_R}  \rho^{(S;0)}_{N_R,M_R} 
(W^{(|M_R|)})^\dagger_{M_R 0_{TL} 0_R;0_S0_{TL}M_R}\\\nonumber
&&+\sum_{{N_S,N_{TL}}\atop{|N_S|+|N_{TL}|=|I_S|-|N_R|}}
\sum_{{I_S,J_S}\atop{|N_R|<|I_S|=|J_S|< N}}
W^{(|I_S|)}_{N_SN_{TL}N_R;I_S 0_{TL} 0_R}  \rho^{(S;0)}_{I_S,J_S} 
(W^{(|J_S|)})^\dagger_{J_S 0_{TL} 0_R;N_SN_{TL}M_R}\\\nonumber
&&+
\sum_{{N_S,N_{TL}}\atop{|N_S|+|N_{TL}| = N-|N_R|}}
W^{(N)}_{N_SN_{TL}N_R;1_R 0_{TL} 0_R}  \rho^{(S;0)}_{1_R,1_R} 
(W^{(N)})^\dagger_{1_R 0_{TL} 0_R;N_SN_{TL}M_R}\,.\\\nonumber
\end{eqnarray}
Thus, among the elements of the 0-order coherence matrix, only two elements can not be perfectly transferred, namely $\rho^{(S;0)}_{0_R,0_R}$ and 
$\rho^{(S;0)}_{1_R,1_R}$, {at that} the normalization
(\ref{norm0})  {becomes} equivalent to
\begin{eqnarray}\label{SR}
\rho^{(S;0)}_{0_R,0_R}+\rho^{(S;0)}_{1_R,1_R}= \rho^{(R;0)}_{0_R,0_R}+\rho^{(R;0)}_{1_R,1_R},
\end{eqnarray}
{ i.e., the sum of the elements from 0- and $N$-excitation blocks is conserved.}

{However, there is a way to overcome this obstacle {and make} the perfect transfer of all elements of the 0-order coherence matrix.
First, we  require 
\begin{eqnarray}\label{SR1}
\rho^{(R;0)}_{1_R,1_R} = \rho^{(S;0)}_{0_R,0_R}.
\end{eqnarray}
Then Eq.~(\ref{SR}) yields
\begin{eqnarray}\label{SR2}
\rho^{(R;0)}_{0_R,0_R} = \rho^{(S;0)}_{1_R,1_R}.
\end{eqnarray}
Now apply the unitary transformation to the receiver (not to the extended receiver) which exchanges the first and last rows (as well as the first and last columns) of the density matrix. Of course, this transformation does not commute with the $z$-projection  of the total spin moment, and therefore it exchanges some elements between 
$n$- and $-n$-order coherence matrices { ($n=1,\dots,N-1$)}. 
This problem does not appears if these rows and columns are zero, i.e., the sender's initial density matrix has the following block-diagonal form:
\begin{eqnarray}\label{Block}
\rho^{(S)}(0)=\left(
\begin{array}{c|c|c}
\rho^{(S;0)}_{0_S,0_S} & 0_{2^N-2} &0\cr
\hline
0_{2^N-2}^T & {\mbox{{\Large{$\tilde \rho^{(S)}$}}}}& 0_{2^N-2}^T\cr
\hline
0&0_{2^N-2} &\rho^{(S;0)}_{1_S,1_S}
\end{array}
\right),
\end{eqnarray}
where $0_{2^N-2}$ and $0^T_{2^N-2}$  are, respectively,  the row and column of $2^N-2$ zeros, and $\tilde \rho^{(S)}$ collects elements of the density matrix in the subspace of states with 1, 2, $\dots$, $N-1$ excitations.
Otherwise, if $\rho^{(S)}(0)$ is a full matrix, one still can restore the structure of the higher-order coherence matrices by applying to the  extended receiver a  unitary transformation which combines the elements of $\pm n$-order coherence matrices with fixed $n$. 
Such transformation does not satisfy the commutation condition  (\ref{st}); we leave this problem beyond the scope of the present work.
}

{
\subsubsection{Restoring of nondiagonal  elements of the 0-order coherence matrix}
\label{Section:0orderRestoring}
Along with the elements of the  higher order  coherence matrix, the nondiagonal elements of 0-order coherence matrix can also be restored, as was shown in \cite{Z_2018} for the two-qubit state. The nondiagonal elements are described by Eq.~(\ref{rhoNMR0cC})  with $N_R\neq M_R$. Restoring these elements reduces to the 
definition of $\lambda^{(0)}_{N_R,M_R}$,
\begin{eqnarray}\label{lam0}
\lambda^{(0)}_{N_R,M_R}=
W^{(|N_R|)}_{0_S0_{TL}N_R;N_R 0_{TL} 0_R} (W^{(|N_R|)})^\dagger_{M_R0_{TL}0_R;0_S 0_{TL} M_R}, \qquad N_R\neq M_R
\end{eqnarray}
and set of constraints for the parameters $\varphi$:
\begin{eqnarray}\label{lam02}
0&=&\sum_{{N_S,N_{TL}}\atop{|N_S|+|N_{TL}|=|I_S|-|N_R|}}W^{(|I_S|)}_{N_SN_{TL}N_R;I_S 0_{TL} 0_R} 
(W^{(|J_S|)})^\dagger_{J_S 0_{TL} 0_R;N_SN_{TL}M_R},\nonumber\\ &&|N_R|=|M_R|<|I_S|=|J_S|<N,
\\\label{lam03}
0&=&\sum_{{N_S,N_{TL}}\atop{|N_S|+|N_{TL}| = N-|N_R|}}
W^{(N)}_{N_SN_{TL}N_R;1_R 0_{TL} 0_R}  
(W^{(N)})^\dagger_{1_R 0_{TL} 0_R;N_SN_{TL}M_R}=0,\quad |N_R|=|M_R|
\end{eqnarray}
{
where at least one of the conditions  $M_R\neq N_R$ or $I_S\neq J_S$ holds. These constraints eliminate nondiagonal elements $\rho^{(S;0)}_{I_S,J_S}$ ($I_S\neq J_S$) from the right hand side (r.h.s) of Eqs.~(\ref{rhoNMR0cC})  with $N_R=M_R$ and remove extra terms from eqs.~(\ref{rhoNMR0cC})  with $N_R\neq M_R$ to provide Eq.~(\ref{lam0}). The diagonal part of system~(\ref{rhoNMR0cC}) with $M_R=N_R$ takes the form
\begin{eqnarray}\label{rhoNMR0cC2}
\rho^{(R;0)}_{N_R,N_R} &=&\\\nonumber
&&
|W^{(|N_R|)}_{0_S0_{TL}N_R;N_R 0_{TL} 0_R}|^2  \rho^{(S;0)}_{N_R,N_R} 
+\sum_{{N_S,N_{TL}}\atop{|N_S|+|N_{TL}| = N-|N_R|}}
|W^{(N)}_{N_SN_{TL}N_R;1_R 0_{TL} 0_R}|^2  \rho^{(S;0)}_{1_R,1_R}\\\nonumber
&&+\sum_{{N_S,N_{TL}}\atop{|N_S|+|N_{TL}|=|I_S|-|N_R|}}
\sum_{{I_S}\atop{|N_R|<|I_S|<N}}
|W^{(|I_S|)}_{N_SN_{TL}N_R;I_S 0_{TL} 0_R}|^2  \rho^{(S;0)}_{I_S,I_S}.\nonumber
\end{eqnarray}
}
We also have to eliminate nondiagonal elements from (\ref{rhoNMR0dC}), imposing one more constraint for $\varphi$:
\begin{eqnarray}\label{lam04}
&&\sum_{{N_S,N_{TL}}\atop{|N_S|+|N_{TL}|=|I_S|}}
W^{(|I_S|)}_{N_SN_{TL}0_R;I_S 0_{TL} 0_R}  
(W^{(|J_S|)})^\dagger_{J_S 0_{TL} 0_R;N_S N_{TL}0_R} =0, \\\nonumber
&&I_S\neq J_S,\;\;0<|I_S|=|J_S|<N.
\end{eqnarray}
Then Eq.~(\ref{rhoNMR0dC}) reduces to
\begin{eqnarray}\label{rhoNMR0dC2}
\rho^{(R;0)}_{0_R,0_R} &=&\rho^{(S;0)}_{0_R,0_R}
\\\nonumber
&&+
\sum_{{N_S,N_{TL}}\atop{|N_S|+|N_{TL}|=|I_S|}}
\sum_{{I_S}\atop{0<|I_S|<N}}
W^{(|I_S|)}_{N_SN_{TL}0_R;I_S 0_{TL} 0_R}  \rho^{(S;0)}_{I_S,I_S} 
(W^{(|I_S|)})^\dagger_{I_S 0_{TL} 0_R;N_S N_{TL}0_R}
\\\nonumber
&&+
\sum_{{N_S,N_{TL}}\atop{|N_S|+|N_{TL}| = N}}
|W^{(N)}_{N_SN_{TL}0_R;1_R 0_{TL} 0_R} |^2 \rho^{(S;0)}_{1_R,1_R} 
.
\end{eqnarray}
Now only diagonal elements of $\rho^{(R;0)}$, described by (\ref{rhoNMR0cC2}), can satisfy the condition of perfect transfer given by Eq.~(\ref{rhoNMR0bz}) with $M_R=N_R$ { and $0<|N_R|<N$}.  

Thus, equations (\ref{lam02})--(\ref{lam04}) have to be added to Eqs.~(\ref{kconstr1}) and (\ref{kconstr2})  in constructing the optimizing unitary transformation of the extended receiver $U_{opt}$.
}{Constraint (\ref{SR1}) is also applicable in this case.} 

\subsection{State-restoring protocol}
Finally, the general protocol for state restoring is the following:
\begin{enumerate}
\item
Create the initial state $\rho(0)$ of the form  (\ref{stst}), (\ref{incon}),
where $\rho^{(S;n)}(0)$ ($|n|>0$) are arbitrary matrices, 
the elements of $\rho^{(S;0)}$ satisfy (\ref{rhoNMR0bz}) and (\ref{SR}) (and perhaps (\ref{SR1}) with the initial state in form (\ref{Block})), 
and $\rho^{(TL,RR)}(0)$ is
\begin{eqnarray}\label{TLR}
\rho^{(TL,R)}={\mbox{diag}}(1,0,0,\dots,0).
\end{eqnarray}
\item
Run evolution of the density matrix up to the time instant  $t_0$.
\item
At the time instant $t=t_0$, apply to the extended  receiver the universally optimal transformation  $U_{opt}$, constructed in Secs.~\ref{Section:0orderNorm} and~\ref{Section:0orderRestoring}.
\item
Construct the matrix  $\rho^{(R)}$ by tracing  the obtained density matrix over the subsystems   $TL$ and  $S$.  
\item
If necessary, apply the unitary transformation exchanging the matrix elements $\rho^{(R;0)}_{0_R,0_R}$ and $\rho^{(R;0)}_{1_R,1_R}$  to completely restore the 0-order coherence matrix. This step is applicable  if~(\ref{SR1}) was used to construct $\rho^{(S;0)}(0)$ and the block-diagonal structure (\ref{Block}) of the sender's initial state $\rho^{(S)}(0)$. 
\end{enumerate}

It is important to emphasize that both $U_{opt}$ and $\rho^{(S;0)} $ are universal objects associated with a particular quantum system and time instant $t_0$ for the receiver's  state registration. {Once constructed,} they can be used for transferring and  restoring {\it any} allowed  higher order coherence matrices. Thus, 0-order coherence matrix serves as a core for transferring the parameters encoded into the higher order coherence matrices. It is also important that equations (\ref{rhoNMR0bz}) describing the perfect transfer  of some elements of $\rho^{(S;0)}$, include the parameters $\varphi$, which are fixed in  constructing the optimized transformation $U_{opt}$.  Therefore the matrix $\rho^{(S;0)}$ depends on the  optimization transformation $U_{opt}$ and must be constructed after fixing parameters $\varphi$.

{
In context of quantum control, a general notion of a Kraus map which is simultaneously optimal for all initial states for a given quantum control problem was  introduced in~\cite{WuPe2007} and called as universally optimal Kraus maps \cite{BoPe2020}. Similarly, we call the unitary transformation $U_{opt}$ and induced by it Kraus map as universally optimal for state recovering since they are optimal for recovering simultaneously all non-zero order coherence matrices.
}

\section{Example: 2-qubit state restoring }
\label{Section:example}

\label{Section:general}
In this section  we consider an example of two-qubit state restoring. 
Now $N=2$ and three multi-quantum coherence matrices have to be considered. Restoring of the non-diagonal part of the 2-qubit transferred matrix was proposed in Ref.~\cite{Z_2018}. {Considering 2-qubit sender and receiver states we  explore other perspectives of 2-qubit state transfer and restoring which were missed in  Ref.\cite{Z_2018}.}
{For instance,  the structure of the 0-order coherence matrix was not optimized therein.}  Here we represent restoring of the 2- and 1-order coherence matrices, requiring a special structure for
the 0-order coherence matrix { which can be almost perfectly transferred up to 2 diagonal elements which provide the trace-normalization (\ref{norm0}). Then we give  remarks on restoring the nondiagonal part  of the 0-order coherence matrix and, finally, construct the  0-order coherence matrix which can be perfectly transferred along the spin chain up to the trivial exchange of two its diagonal elements, see Eq.(\ref{rhoRS2}).}

{We deal with formulas of Sec.\ref{Section:restoring} written for the 2-qubit sender and receiver. These formulas are similar to those derived in \cite{Z_2018},  therefor  we move  them to  Appendix, Sec.\ref{Section:appendix}.  Results below in this section are obtained using those formulas. }


Let us  consider the dynamics of the spin-1/2 chain governed by the $XX$-Hamiltonian taking into account dipole-dipole interactions  among all nodes: 
\begin{eqnarray}\label{XXall}
H=\sum_{j>i} D_{ij}(I_{i;x}I_{j;x}+I_{i;y}I_{j;y}),
\end{eqnarray}
where $D_{ij}=\gamma^2 \hbar/r_{ij}^3$ is the coupling constant,  $\gamma$ is the gyromagnetic ratio, $\hbar$  is the Planck constant. For the homogeneous chain   $r_{i,i+1}=r$ and therefore the coupling constants between the nearest neighbors are the same.

Following Ref.~\cite{Z_2018}, we consider the chain of 42 nodes with two pairs of adjusted 
coupling constants \cite{ABCVV} and chose the time instant $t_0$ for state registration according to (\ref{t_0}):
\begin{eqnarray}
\delta_1=\delta_{N-1}=0.3005 \delta, \qquad \delta_2=\delta_{N-2}=0.5311 \delta, \qquad \delta t_0 = 58.9826,
\end{eqnarray}
where $\delta$ is the nearest-neighbor coupling constant between inner nodes of the chain. 
We take the extended receiver consisting of four nodes and unitary transformation having the block-diagonal form, see (\ref{U})--(\ref{Uvarphi}),
\begin{eqnarray}
&&U={\mbox{diag}}(1,U^{(1)}(\varphi^{(1)}),U^{(2)}(\varphi^{(2)})),\\\nonumber
&&\varphi=(\varphi^{(1)},\varphi^{(2)}),\qquad \varphi^{(1)} =(\varphi^{(1)}_1,\dots,\varphi^{(1)}_{12}),\qquad \varphi^{(2)} =(\varphi^{(2)}_1,\dots,\varphi^{(2)}_{30}),  
\end{eqnarray}
where $U^{(1)}$ and $U^{(2)}$ are, respectively, $4\times 4$ and $6\times 6$ unitary transformations in the subspaces of one and two excited spins. 

{At this stage we shall give a remark. In the above discussion, the multi-index notations were quite formal. Now we  point on the desirable symmetry between nodes of the sender and receiver.  According to this symmetry, the nodes of the receiver must be counted in the reverse order, i.e.,  the first  and the second nodes of the receiver are  the $N$th and the $(N-1)$th nodes of the chain. Hereafter we follow this remark.}

System (\ref{ExN1constr1}), (\ref{ExN1constr3}) consists of 6 independent complex  equations
for the parameters $\varphi$ of the unitary transformation. 
Solution of this system is not unique. Our purpose is to find such unitary transformation which maximizes the scale factors in formulas (\ref{ExN2cohlam1}), (\ref{ExN1lam1}) and (\ref{ExN1lam2}). {  As an {objective function} $J$ for maximization, we take the sum of the absolute values of these parameters,
{
\begin{eqnarray}
J=|\lambda^{(1)}_{00,01}|+|\lambda^{(1)}_{00,10}|+ 
|\lambda^{(1)}_{01,11}|+|\lambda^{(1)}_{10,11}|+|\lambda^{(2)}_{00,11}|\to\max\,,
\end{eqnarray}
}
{ and perform the rough maximization \cite{Z_2018} which gives
\begin{eqnarray}\label{Z}
J=2.391.
\end{eqnarray}
Then we fix $J$ (\ref{Z}),  find 1000 solutions of the system (\ref{ExN1constr1}), (\ref{ExN1constr3}) and
\begin{eqnarray}\label{ZZ}
|\lambda^{(1)}_{00,01}|+|\lambda^{(1)}_{00,10}|+ 
|\lambda^{(1)}_{01,11}|+|\lambda^{(1)}_{10,11}|+|\lambda^{(2)}_{00,11}| = 2.391,
\end{eqnarray}
 and select the solution which 
corresponds to the maximal value of the minimal absolute values of the scale factors, which we denote by  $|\lambda_{min}|$. In our case, the role of $\lambda_{min}$ in the above 1000 solutions is played by either $\lambda_{01,11}$ or $\lambda_{10,11}$ with the maximal by absolute value is  $|\lambda_{01,11}| =0,315 $.  The histogram of calculated values of $|\lambda_{min}|$ is given in Fig.\ref{Fig:H}. 
\begin{figure*}[!]
\epsfig{file=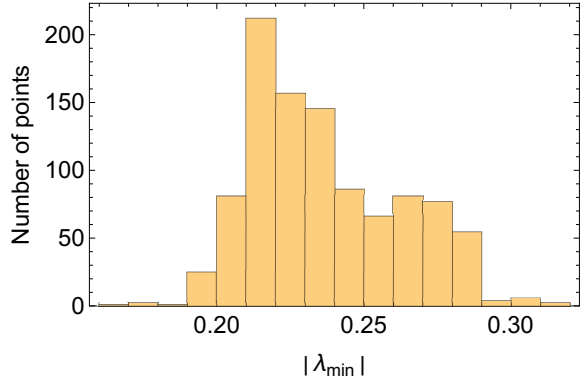,
  scale=0.6
   ,angle=0
} 
\caption{{ Distribution of values of $\lambda_{min}$ for  1000 solutions of system (\ref{ExN1constr1}), (\ref{ExN1constr3}), (\ref{ZZ}).}  
}
 \label{Fig:H} 
\end{figure*}

}}
{ Each solution is found by the Newton method  starting with the random set of initial values for parameters $\varphi$ in the range {$(0,2\pi)$}}. Then we solve system (\ref{rhoNMR0bzN2}) with
\begin{eqnarray}\label{nnn}
(n_1,n_2,m_1,m_2)=(0,1,1,0),\;\;(0,1,0,1),\;\;(1,0,1,0)
\end{eqnarray}
for  $\rho^{(S;0)}_{01,10}$, $\rho^{(S;0)}_{01,01}$, $\rho^{(S;0)}_{10,10}$.
As the result, we obtain $\rho^{(R)}$ {(the scale factors  and the elements $\rho^{(S;0)}_{01,10}$, $\rho^{(S;0)}_{01,01}$, $\rho^{(S;0)}_{10,10}$ are at the appropriate positions in the formula for $\rho^{(R)}$)}
{\small
\begin{eqnarray}\label{EXrho}
&&\rho^{(R)}=\\\nonumber
&&\left(
\begin{array}{cccc}
1- 0.517(1-\rho^{(S)}_{00,00}) & 0.575 e^{-1.812 i }\rho^{(S)}_{00,01}&0.616 e^{3.061i }\rho^{(S)}_{00,10} &  0.548e^{0.257i }\rho^{(S)}_{00,11}\cr
0.575 e^{1.812 i } (\rho^{(S)}_{00,01})^*& 0.214(1-\rho^{(S)}_{00,00}) &0.031 e^{ -2.432i} (1-\rho^{(S)}_{00,00})   &0.315e^{2.070i }\rho^{(S)}_{01,11}\cr
 0.616 e^{-3.061i }(\rho^{(S)}_{00,10})^* &0.031  e^{ 2.432i}(1-\rho^{(S)}_{00,00}) &0.095(1-\rho^{(S)}_{00,00})&0.337e^{ -2.804 i}\rho^{(S)}_{10,11}\cr
0.548e^{-0.257i }(\rho^{(S)}_{00,11})^*&0.315e^{-2.070i }(\rho^{(S)}_{01,11})^*&0.337e^{ 2.804 i}(\rho^{(S)}_{10,11})^*& 0.208(1-\rho^{(S)}_{00,00})
\end{array}
\right)
\end{eqnarray}
}
and the sender's initial state $\rho^{(S)}(0)$ in the form
\begin{eqnarray}\label{rhoSf}
\rho^{(S)}=\left(
\begin{array}{cccc}
\rho^{(S)}_{00,00} & \rho^{(S)}_{00,01}&\rho^{(S)}_{00,10} & \rho^{(S)}_{00,11}\cr
 (\rho^{(S)}_{00,01})^*& 0.214 (1-\rho^{(S)}_{00,00}) &0.031 e^{ -2.432i}(1-\rho^{(S)}_{00,00})&\rho^{(S)}_{01,11}\cr
(\rho^{(S)}_{00,10})^* &0.031 e^{ 2.432i}(1-\rho^{(S)}_{00,00})& 0.095(1-\rho^{(S)}_{00,00})&\rho^{(S)}_{10,11}\cr
 (\rho^{(S)}_{00,11})^*&(\rho^{(S)}_{01,11})^*&(\rho^{(S)}_{10,11})^*& 0.691(1-\rho^{(S)}_{00,00})
\end{array}
\right).
\end{eqnarray}
{In formulas (\ref{EXrho}) and (\ref{rhoSf}), as well as in other density matrices  of this section, we use the basis of two-qubit states $(0,0),\;(0,1),\;(1,0),\; (1,1)$, where $(0,1)$ means the excited edge spin. }
{The non-zero-order coherence matrices include five arbitrary complex parameters: $\rho^{(S)}_{00,01}$, $\rho^{(S)}_{00,10}$,  $\rho^{(S)}_{00,11}$, 
$\rho^{(S)}_{01,11}$, $\rho^{(S)}_{10,11}$. 
{We do not provide the explicit form of $U_{opt}$.}

Further, we consider the constraint (\ref{SR1}) together with the block structure~(\ref{Block}). In our case of two-qubit sender and receiver, the above block structure includes only 0-order coherence matrix, so that there is no higher order coherence matrices. Therefore we disregard $U_{opt}$ (or set $\varphi=0$).  We solve the system (\ref{rhoNMR0bzN2}) with $(n_1,n_2,m_1,m_2)$ from (\ref{nnn})  
together with  Eq.~(\ref{SR1}) {(which now reads $\rho^{(R;0)}_{11;11}= \rho^{(S;0)}_{00;00}$)} for the matrix elements $\rho^{(S;0)}_{01,10}$, $\rho^{(S;0)}_{01,01}$, $\rho^{(S;0)}_{10,10}$, $\rho^{(S;0)}_{00,00}$ taking into account the normalization $\sum_{i,j=0}^1 \rho^{(S;0)}_{ij,ij}=1$. After exchanging the first and last columns and rows in the derived matrix one gets
\begin{eqnarray}\label{rhoRS}
&&\rho^{(R)}(t_0)=\rho^{(S)}(0) =\left(
\begin{array}{cccc}
0.130 & 0&0 & 0\cr
0& 0.487&0.110e^{-1.290i}   &0\cr
0& 0.110e^{1.290i} &0.085\cr
0&0&0&0.298
\end{array}
\right).
\end{eqnarray}
 
Notice that  there are no free parameters in the density matrix (\ref{rhoRS}). 
However, according to Sec.~\ref{Section:0orderRestoring}, the non-diagonal elements of 0-order coherence matrix can carry an arbitrary  parameter. For that, we have to satisfy  condition (\ref{constr0}). 
We perform the rough maximization  similar to maximization in deriving state (\ref{EXrho}) with objective $J=|\lambda^{(0)}_{01,10}|$ to result in the state with single arbitrary parameter $\rho^{(S;0)}_{01,10}$:
 \begin{eqnarray}\label{rhoRS2}
&&\rho^{(R)}(t_0) =\left(
\begin{array}{cccc}
0.006 & 0&0 & 0\cr
0& 0.604&0.560e^{2.393i} \rho^{(S;0)}_{01,10}   &0\cr
0& 0.560e^{-2.393i} (\rho^{(S;0)}_{01,10})^* &0.032\cr
0&0&0&0.358
\end{array}
\right).
\end{eqnarray}
 }

 \section{Conclusions and discussion}
 \label{Section:conclusion}

{We consider the problem of constructing such mixed initial state of the multi-qubit sender which can be transferred to the receiver through the spin chain with minimal and well-characterizable deformation. This deformation is the scaling of the elements of the higher-order coherence matrices (and perhaps non-diagonal elements of the 0-order coherence matrix) of the transfered state and  perfect (or almost perfect) transfer of the 0-order coherence matrix. }

{{ For this purpose,} we consider a  spin-1/2 chain 
consisting of sender  $S$,  transmission line $TL$ and  receiver $R$, where  the sender and the receiver have the same dimension. The system's  dynamics is governed by Hamiltonian preserving the number of excitation in the system. As was  shown in \cite{FZ_2017}, such dynamics  prevents  multi-quantum coherence matrices from mixing.  However, mixing elements inside each coherence matrix had remained an open problem. Partially this problem was resolved in \cite{Z_2018} for a 2-qubit sender and receiver {with the ground initial state of $TL\cup R$,} {where a protocol for restoring non-diagonal elements of the transfered matrix via special unitary transformation of the extended receiver was proposed. But diagonal elements have special properties and can not be optimized in this way.}  

Here we generalize the protocol of Ref.~\cite{Z_2018} to the multi-qubit sender and receiver { of equal dimensions and also study the problem of optimization of the 0-order coherence matrix.} The state of the receiver must be registered at a certain time instant $t_0$. We fix $t_0$ as a time instant maximizing the $N$-order coherence intensity \cite{Z_2018} ($N=2$ in example of Sec.\ref{Section:example}) without applying the unitary transformation to the extended receiver.

For the restoring purpose, { the mentioned above}  extended receiver is introduced into the chain. It includes the receiver as a subsystem and serves to handle the receiver's state through a unitary transformation.   The dimension of the extended receiver must be large enough so that the appropriate unitary transformation possesses enough number of free parameters $\varphi$ to restore the structure of the  elements of the higher order  coherence matrices and, if needed, to optimize the scale factors and to restore the nondiagonal elements of the 0-order coherence matrix.  We describe in detail the protocol for constructing this optimizing unitary transformation  for the sender and the receiver of arbitrary dimension. This constructed transformation is universally optimal, {i.e., it fixes factors ahead of the restored elements and, ones constructed,} can be used to structurally restore any higher-order coherence matrix of the sender's initial state. 

{ The above unitary transformation is not unique. The  variety of such transformations allows to perform further optimization of state restoring. We perform rough optimization calculating 1000 independent unitary transformations and selecting one of them which results in the maximal sum of the absolute values of all scale factors  with the maximal value of the smallest term in this sum.}

We also find such 0-order coherence matrix that can be  { almost perfectly, up to two diagonal elements whose sum is conserved, transferred to the receiver and use this matrix as a block  of the sender's initial state in the proposed protocol for restoring the  higher-order coherence matrices}.  Further, if the initial state of the $N$-qubit sender is embedded into the up to $N$-excitation space then there is the perfectly transferable 0-order coherence matrix  up to the  exchange of two elements corresponding to the blocks of states with 0 and $N$ excitations.} Furthermore,  if the sender's initial state $\rho^{(S)}(0)$  has the block structure (\ref{Block}), then the complete structural restoring of the transferred matrix is possible. Notice that 
each element of the higher-order coherence matrix transfers a free parameter. 
On the contrary,  
the 0-order coherence matrix 
with perfectly transferred nondiagonal elements
does not carry any free parameter because all its elements are fixed.
However,  we can extend the restoring protocol to the non-diagonal elements 
of the 0-order coherence matrix so that free parameters of the initial sender's state appear in the non-diagonal part of 0-order coherence matrix. 
An example of a two-qubit state restoring is explored in details.  

This work was funded by Russian Federation represented by the Ministry of
Science and Higher Education (grant number 075-15-2020-788).

\section{Appendix: general formulas for 2-qubit state restoring}
\label{Section:appendix}
Let $N=2$ and $k=1$ in formulas (\ref{N2cohlam1})-(\ref{klambda2}).
All multi-indexes associated with the sender and receiver consist of two entries $\{n_1,n_2\}$, $n_i=0,1$, $i=1,2$. 
In particular $1_{R}=1_S=\{1,1\}$, $0_{R}=0_S=\{0,0\}$. 
Eq.~(\ref{N2cohlam1}) yields
\begin{eqnarray}\label{ExN2cohlam1}
\lambda^{(2)}_{11,11}=
(W^{(2)})^\dagger_{11, 0_{TL}, 00;00,0_{TL},11}.
\end{eqnarray}
Eqs.~(\ref{kconstr1}) and~(\ref{kconstr2}) yield two equations 
\begin{eqnarray}\label{ExN1constr1}
W^{(1)}_{00,0_{TL},n_1n_2;i_1i_2, 0_{TL}, 00} =0\quad \textrm{ for }
(n_1,n_2) \neq (i_1,i_2), \;\; n_1+n_2=i_1+i_2=1\
\end{eqnarray}
and four equations
\begin{equation}\label{ExN1constr3}
\sum_{{i_1,i_2,N_{TL}}\atop{|N_{TL}|+i_1+i_2=1}}
W^{(1)}_{i_1i_2, N_{TL},00;n_1n_2, 0_{TL}, 00}  
(W^{(2)})^\dagger_{11, 0_{TL}, 00;i_1i_2, N_{TL}, m_1m_2} =0, \quad |I_S|=|M_R|=1\,.
\end{equation}
In addition, eqs.~(\ref{klambda1}) and~(\ref{klambda2}) yield the following definitions of the scale factors:
\begin{eqnarray}\label{ExN1lam1}
\lambda^{(1)}_{n_1n_2, 11}&=&W^{(1)}_{00,0_{TL}, n_1n_2;n_2n_2, 0_{TL}, 00} W^{(2)}_{11,0_{TL},00;00,0_{TL},11},\qquad n_1+n_2=1\\
\label{ExN1lam2}
\lambda^{(1)}_{00, m_1m_2} &=& (W^{(1)})^\dagger_{m_1m_2, 0_{TL}, 00;00,0_{TL}, m_1m_2},\qquad m_1+m_2=1.
\end{eqnarray}
For the elements $\rho^{(R;0)}_{11,11}$ and $\rho^{(R;0)}_{00,00}$ of the 0-order coherence matrix (\ref{rhoNMR0a}), (\ref{rhoNMR0dC}), we have
\begin{eqnarray}
\rho^{(R;0)}_{11,11}&=& | W^{(2)}_{00,0_{TL},11;11,0_{TL},00}|^2 \rho^{(S;0)}_{11,11},\\\label{R1111}
\rho^{(R;0)}_{00,00}&=& \rho^{(S;0)}_{00,00}+
\sum_{{k_1,k_2,N_{TL}}\atop{k_1+k_2+|N_{TL}|=2}}
|W^{(2)}_{k_1k_2,N_{TL},00;11, 0_{TL},00}|^2  \rho^{(S;0)}_{11,11} \\\nonumber
&&+
\sum_{{k_1,k_2,N_{TL}}\atop{k_1+k_2+|N_{TL}|=1}}
\sum_{{i_1+i_2=}\atop{j_1+j_2=1}}
W^{(1)}_{k_1k_2,N_{TL},00;i_1i_2, 0_{TL},00}  \rho^{(S;0)}_{i_1i_2,j_1j_2} 
(W^{(1)})^\dagger_{j_1j_2, 0_{TL},00;k_1k_2, N_{TL},00}\, .
\end{eqnarray}
The last term in Eq.~(\ref{R1111}) can be simplified using unitarity of $W^{(1)}$ which assumes
\begin{eqnarray}
\sum_{{k1,k_2,l_1,l_2,N_{TL}}\atop{{k_1+k_2+l_1+}\atop{l_2+|N_{TL}|=1}}} W^{(1)}_{k_1k_2,N_{TL},l_1l_2; i_1i_2,0_{TL},00} (W^{(1)})^\dagger_{i_2i_1,0_{TL},00;k_1k_2,N_{TL},l_1l_2} =0,\;\; i_1+i_2=j_1+j_2=1,
\end{eqnarray}
and condition (\ref{ExN1constr1}), so that
\begin{eqnarray}\label{rho00ex}
\rho^{(R;0)}_{00,00}&=& \rho^{(S;0)}_{00,00} 
\\\nonumber
&&+
\sum_{{k_1,k_2,N_{TL}}\atop{k_1+k_2+|N_{TL}|=2}}
|W^{(2)}_{k_1k_2,N_{TL},00;11, 0_{TL},00}|^2  \rho^{(S;0)}_{11,11} 
\\\nonumber
&&+
\sum_{{k_1,k_2,N_{TL}}\atop{k_1+k_2+|N_{TL}|=1}}
\sum_{{i_1,i_2}\atop{i_1+i_2=1}}
|W^{(1)}_{k_1k_2,N_{TL},00;i_1i_2, 0_{TL},00}|^2  \rho^{(S;0)}_{i_1i_2,i_1i_2} .
\end{eqnarray}

Other elements of $\rho^{(R;0)}$ satisfy Eq.~(\ref{rhoNMR0bz}) which now takes the form
\begin{eqnarray}\label{rhoNMR0bzN2}
\rho^{(R;0)}_{n_1n_2,m_1m_2} &=&\rho^{(S;0)}_{n_1n_2,m_1m_2} 
\\\nonumber
&=&
W^{(1)}_{00,0_{TL},n_1n_2;n_1n_2, 0_{TL},00}  \rho^{(S;0)}_{n_1n_2,m_1m_2} 
(W^{(1)})^\dagger_{m_1m_2, 0_{TL},00 ;00,0_{TL},m_1m_2}\\\nonumber
&&+
\sum_{{k_1,k_2,N_{TL}}\atop{k_1+k_2+|N_{TL}|=1}}
W^{(2)}_{k_1k_2,N_{TL},n_1n_2;11, 0_{TL},00}  \rho^{(S;0)}_{11,11} 
(W^{(2)})^\dagger_{11, 0_{TL},00 ;k_1k_2,N_{TL},m_1m_2},\\\nonumber
&&
n_1+n_2=m_1+m_2=1.
\end{eqnarray}
The condition (\ref{SR}) is satisfied in this case, the constrain (\ref{SR1}) can also be imposed.

{
From the structure of Eqs. (\ref{rho00ex}) and (\ref{rhoNMR0bzN2}) it follows that, for restoring the non-diagonal part of $\rho^{(S;0)}$, we  have the equation:
\begin{eqnarray}\label{constr0}
\sum_{{k_1,k_2,N_{TL}}\atop{k_1+k_2+|N_{TL}|=1}}
W^{(2)}_{k_1k_2,N_{TL},01;11, 0_{TL},00}  
(W^{(2)})^\dagger_{11, 0_{TL},00 ;k_1k_2,N_{TL},10}=0
\end{eqnarray}
with the additional scale factor
\begin{eqnarray}
\lambda^{(0)}_{01,10} =W^{(1)}_{00,0_{TL},01;01, 0_{TL},00}   
(W^{(1)})^\dagger_{10, 0_{TL},00 ;00,0_{TL},10}.
\end{eqnarray}
}

\end{document}